\documentclass[pra,reprint,superscriptaddress,showpacs]{revtex4-2}

\usepackage[T1]{fontenc}
\usepackage{booktabs}
\usepackage{mathptmx}
\usepackage{graphics}
\usepackage{multirow}
\usepackage{subfigure}
\usepackage{graphicx}
\usepackage{amsmath}
\usepackage{makecell}
\usepackage[colorlinks,
linkcolor=blue,
anchorcolor=red,
urlcolor=blue,
citecolor=blue]{hyperref}
\usepackage{threeparttable}

\begin{document}
\renewcommand{\thetable}{\Roman{table}}
\title{\textbf{Ta$_{2}$Pd$_{3}$Te$_{8}$$:$ A potential candidate of 1D van der Waals stacked thermoelectric materials
  }}%

\author{Shi Chen}%
\affiliation{School of Physics, Communication and Electronics, Jiangxi Normal University, Nanchang 330022, China }

\author{Aijun Hong}%
	\email[Corresponding author: ]{6312886haj@163.com}
	
	\affiliation{School of Physics, Communication and Electronics, Jiangxi Normal University, Nanchang 330022, China }


	
	\author{Junming Liu}%
	\email[Corresponding author: ]{liujm@nju.edu.cn}
	\affiliation{Laboratory of Solid State Microstructures, Nanjing University, Nanjing 210093, China}

\begin{abstract}
 Discovering new thermoelectric (TE) materials is an eternal goal in the TE field. Excellent TE materials have ranged from 3D stacked to 2D stacked bulk. However, the 1D stacked receives little attention due to the scarcity in quantity. In this work, it is predicted  that 1D van der Waals (vdW) stacked Ta$_{2}$Pd$_{3}$Te$_{8}$ crystal is a compelling candidate for TE applications by combining first-principles calculations with phonon and electron Boltzmann transport equations and molecular dynamics methods. We find that Ta$_{2}$Pd$_{3}$Te$_{8}$ crystal has mechanical, dynamical, and thermal stabilities, and its TE properties are featured by strong anisotropy, high power factor ($PF$) and low lattice thermal conductivity. The results indicate the $ZT$ values of n-type Ta$_{2}$Pd$_{3}$Te$_{8}$ at 900 K along $a$, $b$ and $c$ axes reach 0.48, 0.39 and 0.22, respectively.  We propose that enlarging the bandgap can weaken the bipolar effect and thus significantly increases $ZT$ to 1.11. The findings in the work not only stimulate more theoretical works on 1D vdW stacked TE materials, but also provide valuable information for experimentally improving TE materials.

\end{abstract}
\maketitle

\section{Introduction}
Thermoelectric (TE) materials that can convert thermal energy and electrical energy into each other, have attained tremendous attention, because they support possible strategies to address environmental pollution and energy shortage \cite{r1,r2,r3}.
To find new types of TE materials and optimize already-existent TE materials by novel technologies are two major routes for TE research. However, the improvement of  TE performance is limited by the complexly contradictory relationship,

\begin{equation} \label{ee3}
ZT=\frac{\sigma S^2}{\kappa^e + \kappa^l }T,
\end{equation}
where $\sigma$ and $S$ are the electrical conductivity and Seebeck coefficient, $T$ is the absolute temperature, and $\kappa^e$ and $\kappa^l$ stand for the electronic and lattice contributions to the total thermal conductivity $\kappa$.
To reduce the total thermal conductivity and improve the power factor $PF=\sigma S^2$ are the most direct and effective methods.
Nevertheless, the $S$ is positively correlated with effective mass and carrier density, while the $\sigma$ is exactly the opposite case. This hinders their collective improvements.
Furthermore, the $\kappa^e$  is directly proportional to the $\sigma$ according to the Wiedemann-Franz law, implying the electrical conductivity cannot be increased indefinitely. This poses significant obstacles to regulating TE performance.
Besides, high electrical conductivity materials are often accompanied by high $\kappa^l$. This is because high $\sigma$ implies weak scattering of electron suffering from phonon and thus indicates weak anharmonic vibration that leads to high $\kappa^l$.
In a word, that discovering new type parent TE materials with weak coupling between transport parameters $\sigma$, $S$, $\kappa^l$ and $\kappa^e$ and then making improvements is particularly important.

In recent decades, phonon-liquid electron-crystal Cu$_2$Se \cite{r4}, anisotropic layered SnSe \cite{r5,r6,r7} and Bi$_2$Te$_3$ \cite{r11,r12}, half-Husler alloy NbSbFe \cite{r8,r9}, and PbSe\text{-}based high-entropy alloys \cite{r10} have been experimentally confirmed to be promising TE materials owing to the weak coupling between transport parameters.
Especially, layered stacked crystals SnSe and Bi$_2$Te$_3$ greatly stimulate the enthusiasm for studying the TE properties of single crystal materials.
Usually, the layered stacked materials has a low-symmetry structure.
As known, crystals with the low symmetry typically exhibit anisotropic electrical and thermal transports.
The anisotropy provides the possibility of the decoupling between transport parameters along a specific direction, and thus achieves improved TE performance.

The anisotropy of Bi$_2$Te$_3$ as well as that of SnSe originates from 2D layered stacked cystal structure, and results in the large differences between the transport parameters along different directions.
The in-plane electrical conductivity $\sigma_{\|}$ is greater than that along the out-plane $\sigma_{\bot}$ with the ratio ($\sigma_{\|}/\sigma_{\bot}$) as large as 7 in n-type and 4 in p-type Bi$_2$Te$_3$ \cite{haj2}. The anisotropy in the thermal
conductivity $\kappa_{\|}$/$\kappa_{\bot}$ as high as 2.5 \cite{haj2}. Hence, by leveraging these features, the $ZT$ value of Bi$_2$Te$_3$ single crystal \cite{r3} has been increased to 1.05 at room temperature via doping engineering.

SnSe was first reported in the year of 2014 to have record high $ZT$ values of 2.6 and 2.3 \cite{r5} at about 900 K in the p-type crystal along the \emph{b} and \emph{c} axes in the in-plane, respectively. The anisotropy of SnSe crystal plays a vital role in regulating TE performance. Besides, it seems to support a false impression that the TE performance in the in-plane should be superior to that in the out-plane. However, by means of doping Br at Se site, a higher $ZT$ value of 2.8 \cite{r7} was observed along the unexpected \emph{a} axis in the out-plane of n-type SnSe crystal. The Br substitution leads to the overlapping interlayer charge density exhibiting 3D charge transport, while having slight influence on the thermal transport with low out-plane thermal conductivity termed by 2D phonon transport.

At present, there are a large number of layered stacked materials available for selection.
Thus, researchers have devoted a significant amount of vigor to discovering and improving 2D layered stacked TE materials, and seem to be indifferent to the 1D stacked ones due to its scarcity in quantity.
Recently, a family of the 1D vdW stacked bulk materials termed by channel materials or 1D vdW materials, attract great attention owing to unique properties resulting from the distinct crystal structure and thus being promising applications such as in transistors and optoelectronics \cite{r16,r17}.
For instance, experimentally, 1D vdW Nb$_2$Pd$_3$Se$_8$ displays a transition from indirect to direct band gap \cite{r18} due to structural variation from bulk to single, and field effect transistors based on its nanowires exhist the n-type mobility and I$_{on}$/I$_{off}$ ratio of 32 cm$^{2}$V$^{-1}$s$^{-1}$ and approximately 10$^{4}$ \cite{r19}, respectively. Moreover, the calculated n-type mobility of Ta$_2$Ni$_3$Se$_8$ bulk reaches as high as 264 cm$^{2}$V$^{-1}$s$^{-1}$ \cite{r18}. In fact, carrier mobility is an important parameter for excellent TE materials.

The 1D vdW materials are made up of countless chain-like nanowires that are built through covalent/ionic bonds, and have weak vdW forces between the nanowires. Thus, a few chain-like nanowires could be easily obtained by directly exfoliating from the bulk, like 2D stacked materials.
Furthermoe, the 1D vdW stacked materials, compared with the 2D stacked ones, further dissociate the coupling between electrical and thermal transports.
The characteristic of dangling-bond-free in the interstices between chains mitigates the deterioration in electrical transport caused by the scattering around the dangling bonds.
During phonon transmission, there will be additional scattering in two dimensions of the 1D vdW stacked materials. Obviously, the crystal structures of the 1D vdW stacked materials are  conducive to achieving decoupling of TE parameters, thereby improving TE performance.

In this work, we focus on TE properties of Ta$_{2}$Pd$_{3}$Te$_{8}$ crystal.
Herein, the first-principles method incorporating the Boltzmann transport theory and molecular dynamics methods were employed to systematically investigate crystal, electronic and  phononic structures, and thermal and electronic transport properties of Ta$_{2}$Pd$_{3}$Te$_{8}$ crystal.

\section{Methodology}
The first-principles calculations based on density functional theory (DFT) were performed in Vienna ab initio simulation package (VASP) \cite{r22,r23} and WIEN2k code \cite{r20}.
 First, the crystal structure was fully relaxed with the electronic convergence of 10$^{-6}$ eV, ionic convergence of 1 meV/\AA, ~and \textbf{k}-mesh of 6$\times$4$\times$15 points by using the VASP code.
 Again, the internal structural parameters of the preliminary optimized structure were relaxed with the energy convergence of 0.1 mRy and force convergence of 1 mRy/a.u. by using WIEN2k code  that adopts the all-electron full-potential approach based on linearized augmented plane wave (LAPW).
 The exchange-correlation effects were described by the Perdew-Burke Ernzerhof (PBE) functional \cite{r24,r25} modified by Grimme's DFT-D3 vdW corrections (PBE-D3).
 In the static self-consistent calculation, the modified Becke-Johnson exchange potential (mBJ) \cite{r21} was integrated into the PBE-D3 correlation potential (PBE-D3-mBJ) in order to obtain the band gap with an accuracy comparable to very expensive GW calculations.
 Noted that the vdW correction  was neglected in the calculations on the phonon because of limited computing resources and the slight influence of weaken vdW on the atomic vibrations.

Second, elastic constants, phonon spectrum, and molecular dynamics were employed to verify  mechanical, dynamical, and thermal stabilities. The elastic constants were obtained by applying finite distortions to the lattice and relying on the strain-stress relationship, so that related sound velocity, Debye temperature and so on were computed with the help of VASPKIT code \cite{s1}. The phonon DOS and band structure were calculated by the Phonopy package combined with the VASP via the finite-displacement approach. The molecular dynamics calculations were carried out in the VASP code. the canonical ensemble (NVT) was adopted to simulate the free energy as a function.

Finally, the electronic transport properties were calculated by solving the electron Boltzmann transport equation (EBTE) based on the constant relaxation time approximation, as implemented within the BoltzTraP code \cite{rr22}. We adopted a higher \textbf{k}-mesh of $33\times23\times100$ for self-consistent calculations so as to obtain accurate transport coefficients.
The relaxation times as a function of temperatures were computed according to the mobility formula based on the deformation potential (DP) theory.
The second-order and third-order interatomic force constants (IFCs) were obtain by using compressive sensing lattice dynamics (CSLD) method \cite{rr23,rr24} where a $2\times2\times2$ (208 atoms) surpercell and a \textbf{k}-mesh of $1\times1\times4$ were employed. 60 supercells with random displacement of 0.01 \AA ~and 30 supercells with random displacement of 0.05 \AA ~were modeled to calculate the second-order and third-order IFCs, respectively.
Therefore, the phonon transport properties were calculated by solving the phonon Boltzmann transport equation (PBTE)  implemented in the ShengBTE code \cite{rr25,rr26} and with a $8\times6\times24$ \textbf{q}-mesh

\section{Results and discussion}
\subsection{Structural stability}
\begin{figure*}[htp]
\centering
\includegraphics[width=1.4\columnwidth]{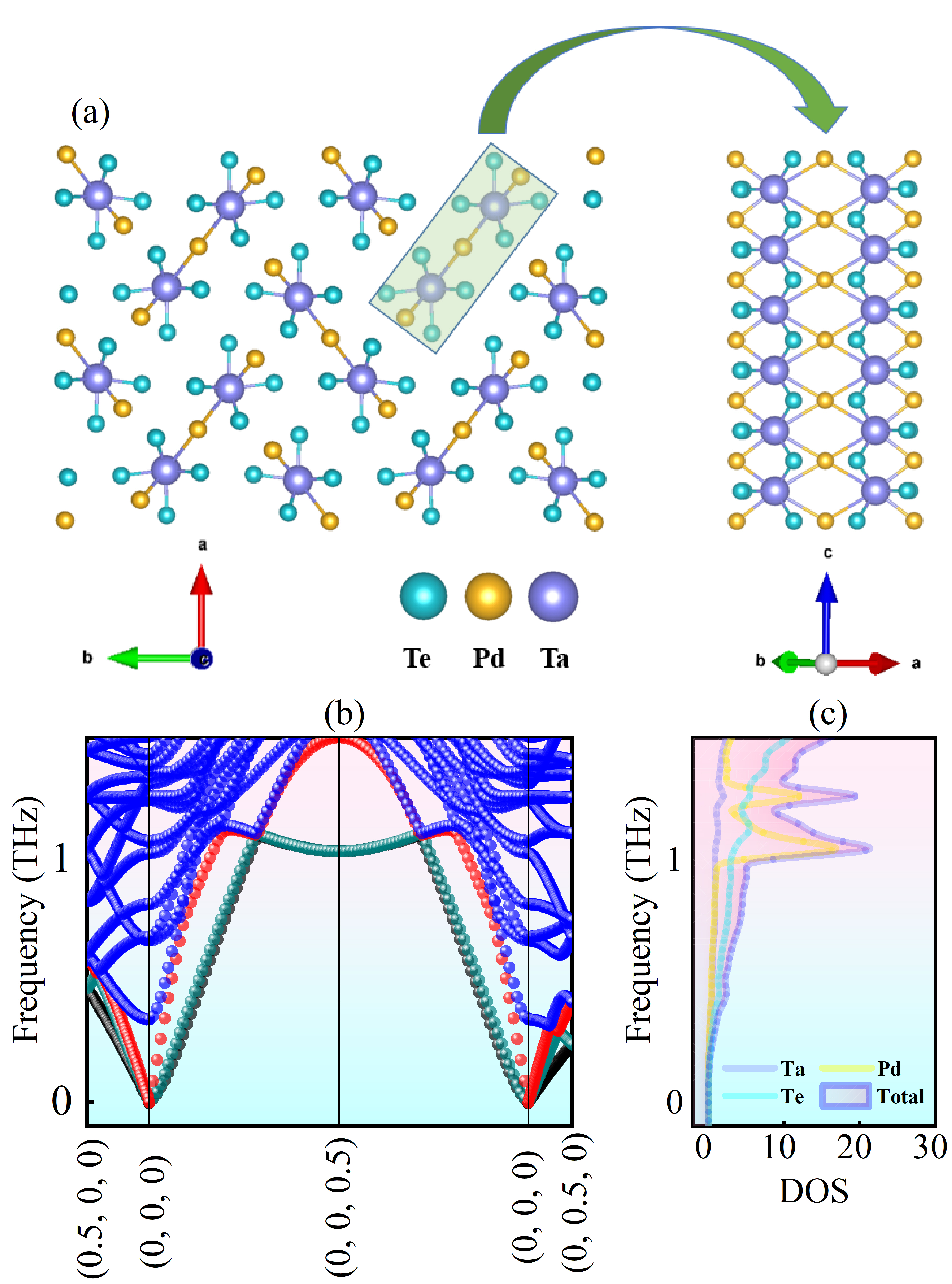}
	\caption{(Color online)
(a) Schematic crystal structure of Ta$_{2}$Pd$_{3}$Te$_{8}$: The left is a supercell view projected along the $c$ axis and the marked area is top view of a chain ribbon, and the right is the side view of the chain ribbon. (b) Phonon dispersion: blue, red, black and cyan stand for optical branches, LA branch and two TA branches, respectively. (c)Total and atom-resolved phonon densities of states.
    \label{Fig001} }
\end{figure*}
Fig. \hyperref[Fig001]{\ref*{Fig001}}(a) depicts the Ta$_{2}$Pd$_{3}$Te$_{8}$ crystal is constituted by a great number of repeating chain ribbons that consist of two Ta-centered pseudo-square cylinders capped by Pd and Te atoms and jointed by one Pd atom. The chain ribbons form a zigzag shape in the \emph{ab} plane and extend along the \emph{c} axis.
\begin{table}[h]  
 \caption{\label{t1} The calculated elastic constants of Ta$_{2}$Pd$_{3}$Te$_{8}$ in units of GPa.}
  \centering  
  \begin{tabular}{ccccccccc}  
    \toprule
   $C_{11}$ & $C_{22}$ & $C_{33}$ & $C_{44}$ & $C_{55}$ & $C_{66}$ & $C_{12}$ & $C_{13}$ & $C_{23}$ \\
    \midrule
    53.9 & 63.4 & 87.8 &  10.6 &  10.2  &  23.8 &  21.7 & 17.7&18.3\\
     \bottomrule
  \end{tabular}
  \label{tab1}  
\end{table}

Structural stability is the most fundamental and important property for TE materials operating at high temperatures. Herein, we employed the elastic constants ($ C_{ ij }$) to confirm the mechanical stability of Ta$_{2}$Pd$_{3}$Te$_{8}$ bulk. The structure of Ta$_{2}$Pd$_{3}$Se$_{8}$ belongs to orthorhombic phase (space group \emph{$Pbam$}, No. 55). Hence, there are nine independent constants in its elastic matrix
\begin{equation}  \label{ee1}
\mathbf{C_{ortho}} =
\begin{pmatrix}
    C_{11}&C_{12}&C_{13}&&& \\
    &C_{22}&C_{23}&&& \\
    &&C_{33}&&& \\
    &&& C_{44}& \\
    &&&& C_{55}& \\
    &&&&&C_{66}\\
\end{pmatrix}
,
\end{equation}
and its mechanical stability criteria are govern by \cite{rr27}
\begin{gather} \label{ee2}
C_{11}>0,C_{44}>0,C_{55}>0,C_{66}>0,\notag \\
C_{11}C_{22}C_{33}+2C_{12}C_{13}C_{23}-C_{11}C_{23}^2-C_{22}C_{13}^2-C_{33}C_{12}^2 > 0,\notag \\
C_{11}C_{22}>C_{12}^2.
\end{gather}
Notably, the nine elastic constants in Table \hyperref[t1]{\ref*{t1}} meet the above criteria well, implying mechanical stability.
Phonon dispersion relations can reflect dynamic characteristics and thus are used to demonstrate the dynamic stability of crystal structures. The phonon dispersion curves of Ta$_{2}$Pd$_{3}$Te$_{8}$ in Fig. \hyperref[Fig001]{\ref*{Fig001}}(b) indicate there are no imaginary phonon frequencies implying dynamic stability.
\begin{figure*}[t]
\centering
\includegraphics[width=1.7\columnwidth]{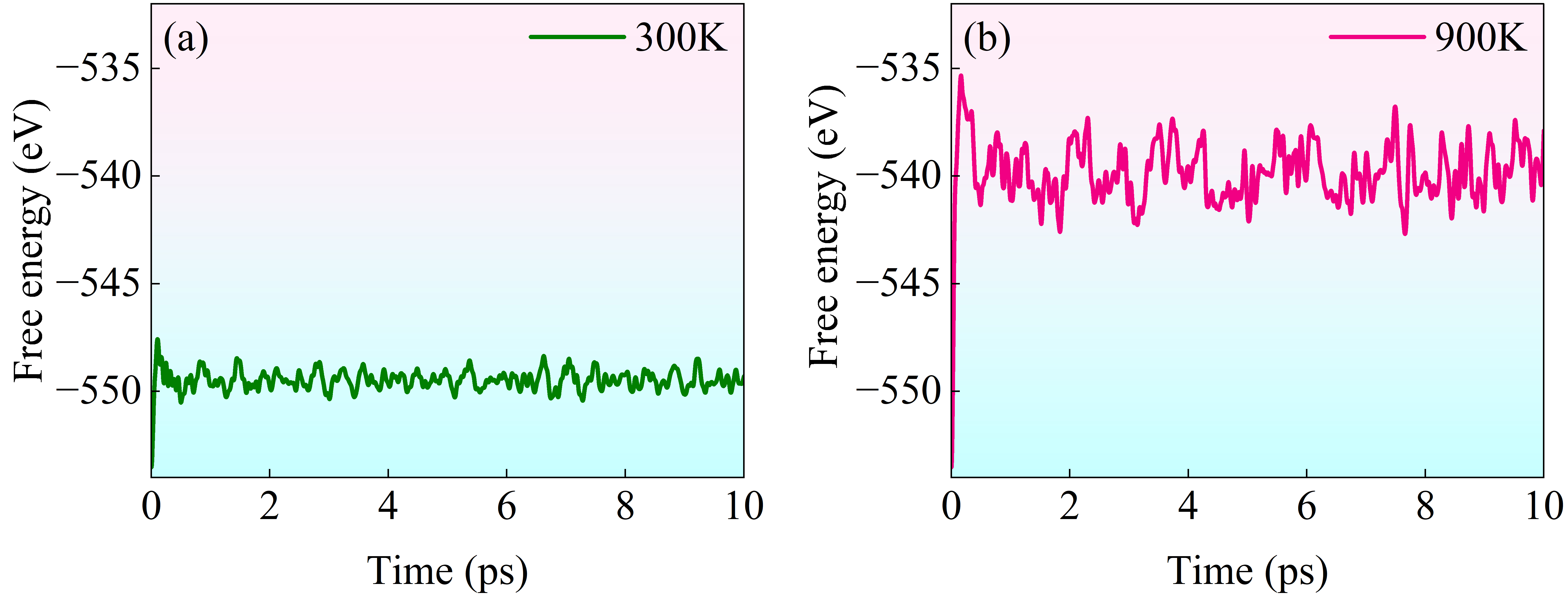}
	\caption{(Color online)
Free energy as a function of time at 300 and 900 K [(a) and (b)]. Energy fluctuations is around a fixed value exhibiting thermal stability of Ta$_{2}$Pd$_{3}$Te$_{8}$ crystal, although the fluctuation at 900 K is larger than that at 300 K.
    \label{Fig002} }
\end{figure*}

A primitive cell involves 26 atoms composed of 4 Ta, 6 Pd and 16 Te atoms, therefore, there are 78 phonon dispersion curves. The highest frequency of the optical bands is only 6.25 THz that is relatively low compared with good TE materials such as half-Heusler alloy FeNbSb, implying that Ta$_{2}$Pd$_{3}$Te$_{8}$ has a possible low lattice thermal conductivity.
The acoustic bands containing two transverse acoustic (TA) branches and one longitudinal acoustic (LA) branch have a powerful role on the thermal transport, and have strong anisotropy as a result of low-symmetry structure.
The highest frequency of the LA modes is 1.5 THz located at the site (0,0,0.5), which is higher than that of TA mode (1.1 THz). The highest frequencies of acoustic modes along the (0.5,0,0.0) and (0,0.5,0) directions are lower than 1 THz. Thus, the thermal transport along the \emph{a} and \emph{b} axes should be inferior to that along the \emph{c} axis, which is further confirmed by subsequent lattice thermal conductivity results.

Normally, in many materials, two TA branches are degenerate, however, they are non-degenerate in Ta$_{2}$Pd$_{3}$Te$_{8}$ crystal. Besides, LA branch and one TA branch tend to degenerate along the \emph{a} and \emph{b} axes, and the two TA bands tend to degenerate along the \emph{c} axis in Ta$_{2}$Pd$_{3}$Te$_{8}$. These are ascribed to low symmetry and one-dimensional chain like structure. The avoided crossings occur not only between the acoustic bands, but also between acoustic and optical bands, implying the strong phonon-phonon interactions. The phononic density of states (PDOS) plot in Fig. \hyperref[Fig001]{\ref*{Fig001}}(c) shows that the vibration of Te atoms contributes the most to the total DOS in the low-frequency range, and the contributions of Ta and Pd atoms are comparable. Above 1 Hz, the contribution of Pd atoms becomes the largest, resulting from the enhanced hybridization between Pd, Ta and Se atoms.

Subsequently, ab initio molecular dynamics (AIMD) simulations with the canonical
ensemble (NVT) are performed in VASP for confirming the thermal stability of Ta$_{2}$Pd$_{3}$Te$_{8}$ bulk. The free energy fluctuations as a function of time at two temperatures (300 K and 900 K) exhibit the free energy fluctuates around a fixed energy after a short heating time (see Fig. \hyperref[Fig002]{\ref*{Fig002}}). The fluctuates at 900 K are more serious than that at 300 K due to thermal fluctuation effect. The geometric structure and atomic coordinates have not undergone significant changes, therefore,  Ta$_{2}$Pd$_{3}$Te$_{8}$ bulk is thermal stable in the temperature range of 300-900 K.

\subsection{Thermoelectric property}

\begin{figure*}[t]
\centering
\includegraphics[width=1.7\columnwidth]{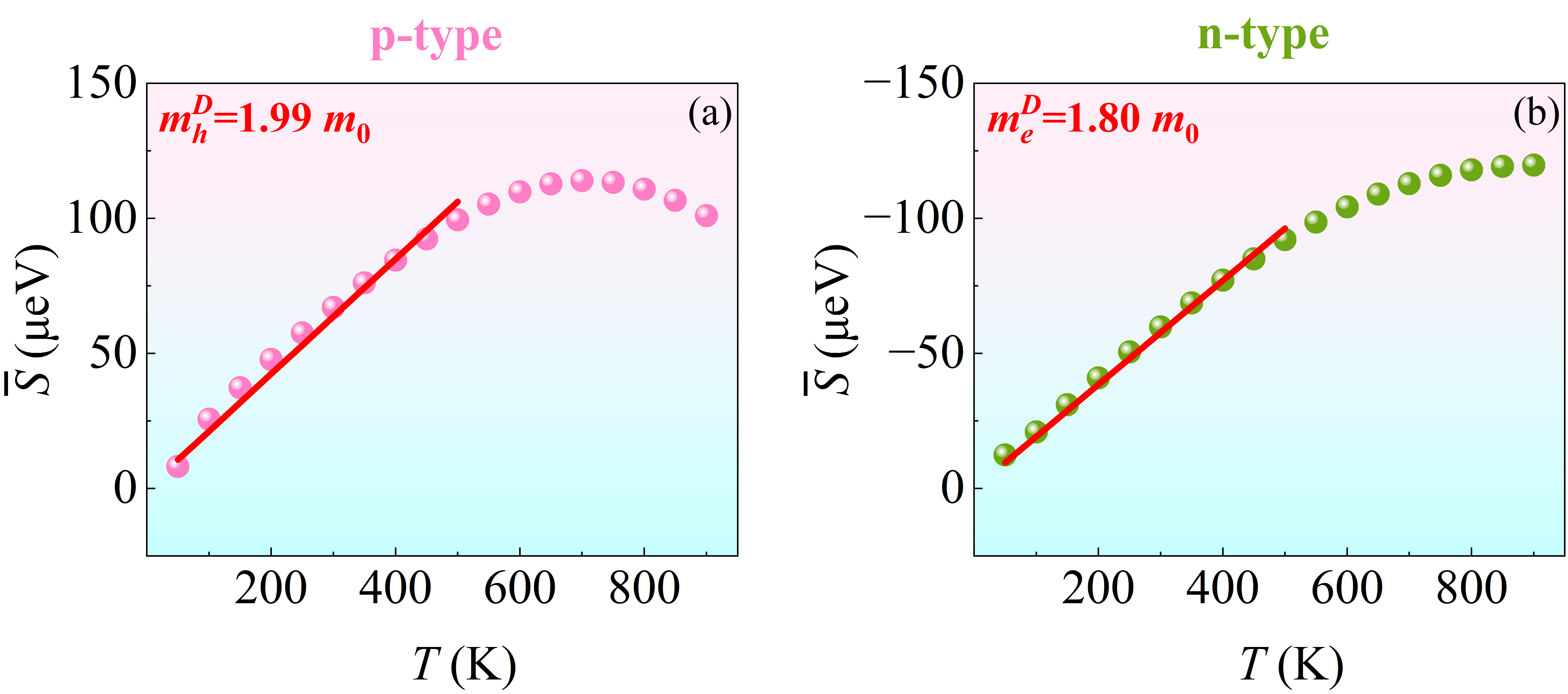}
	\caption{(Color online) The average value $\overline{S}$ of $S_a$, $S_b$ and $S_c$ of the p-type (a) and n-type (b) Ta$_{2}$Pd$_{3}$Te$_{8}$ as a function of temperature. The carrier densities of n- and p-type are fixed at $5\times10^{20} cm^{-3}$. The DOS effective masses are obtained by fitting the $\overline{S}$ values from 300 to 500 K.
    \label{Fig003} }
\end{figure*}

In fact, the four parameters $\sigma$, $S$ and $\kappa^e$ and $\kappa^l$ in Eq.~\ref{ee3} are 3$\times$3  tensors as a function of chemical potential $\varepsilon$ and $T$ within the framework of the classic Boltzmann theory. Therefore, anisotropic $ZT$ for  the single crystal can be written as
\begin{equation} \label{ee4}
ZT_{ij}=\frac{\sigma_{ij} S_{ij}^2}{\kappa^e_{ij} + \kappa^l_{ij} }T, ~~(i, j=1, 2, 3).
\end{equation}
Moreover, the isotropous $\overline{ZT}$ for the polycrystal is defined as
\begin{equation} \label{ee5}
\overline{ZT}=\frac{\overline{\sigma}~\overline{S}^2}{\overline{\kappa^e} + \overline{\kappa^l} }T,\\
\end{equation}
where $\overline{\sigma}$, $\overline{S}$, $\overline{\kappa^e}$ and $\overline{\kappa^l}$ can be approximated as the average values of $\sigma_{ii}$, $S_{ii}$, $\kappa^e_{ii}$ and $\kappa^l_{ii}$ along three principle directions ($i=1, 2, 3$), respectively,
\begin{equation} \label{ee6}
\begin{aligned}
&\overline{\sigma}~=\frac{\sigma_{11} + \sigma_{22} + \sigma_{33}}{3},\\
&\overline{S}~~=\frac{S_{11} + S_{22} + S_{33}}{3},\\
&\overline{\kappa^e} =\frac{\kappa^{e}_{11} + \kappa^{e}_{22} + \kappa^{e}_{33}}{3},\\
&\overline{\kappa^l} =\frac{\kappa^l_{11} + \kappa^l_{22} + \kappa^l_{33}}{3},
\end{aligned}
\end{equation}
For orthorhombic crystal,  $\sigma_{11}$,  $\sigma_{22}$ and  $\sigma_{33}$ can be written as $\sigma_{a}$,  $\sigma_{b}$ and  $\sigma_{c}$, and $S_{ii}$, $\kappa^e_{ii}$ and $\kappa^l_{ii}$ can also be marked in this way.

Under the constant relaxation time ($\tau$) approximation, the $\sigma_{ij}/\tau$, $\kappa^e_{ij}/\tau$ and $S_{ij}$ (independent of $\tau$) are obtained by solving the electron Boltzmann equation. Therefore, obtaining the relaxation time is crucial for calculating electrical transport parameters.
In the frame of the DP theory \cite{rr28}, one can obtain the carrier mobility expression for the isotropy crystals without polar and degeneracy
\begin{equation}\label{ee7}
  \mu_{h(e)}=\frac{2\sqrt{2\pi}q\hbar^{4}\rho v^2}{3(k_{B}T)^{3/2}m^{3/2}_{h(e)}m_{h(e)}\lambda_{h(e)}^2},
\end{equation}
where $k_{B}$, $q$ and $\hbar$ represent the Boltzmae constant, the electronic charge and the reduced Planck constant, and $\rho$ and $v$ stand for the mass density and the LA velocity, and $m_{h(e)}$ and $\lambda_{h(e)}$ are the effective mass and DP constant for the hole(electron) carrier. Typically, $m_{h(e)}$ and $\lambda_{h(e)}$ are assumed to be the effective mass and DP constant at valence band maximum (VBM) or the conduction band minimum (CBM).

\begin{figure*}[htp]
\centering
\includegraphics[width=2\columnwidth]{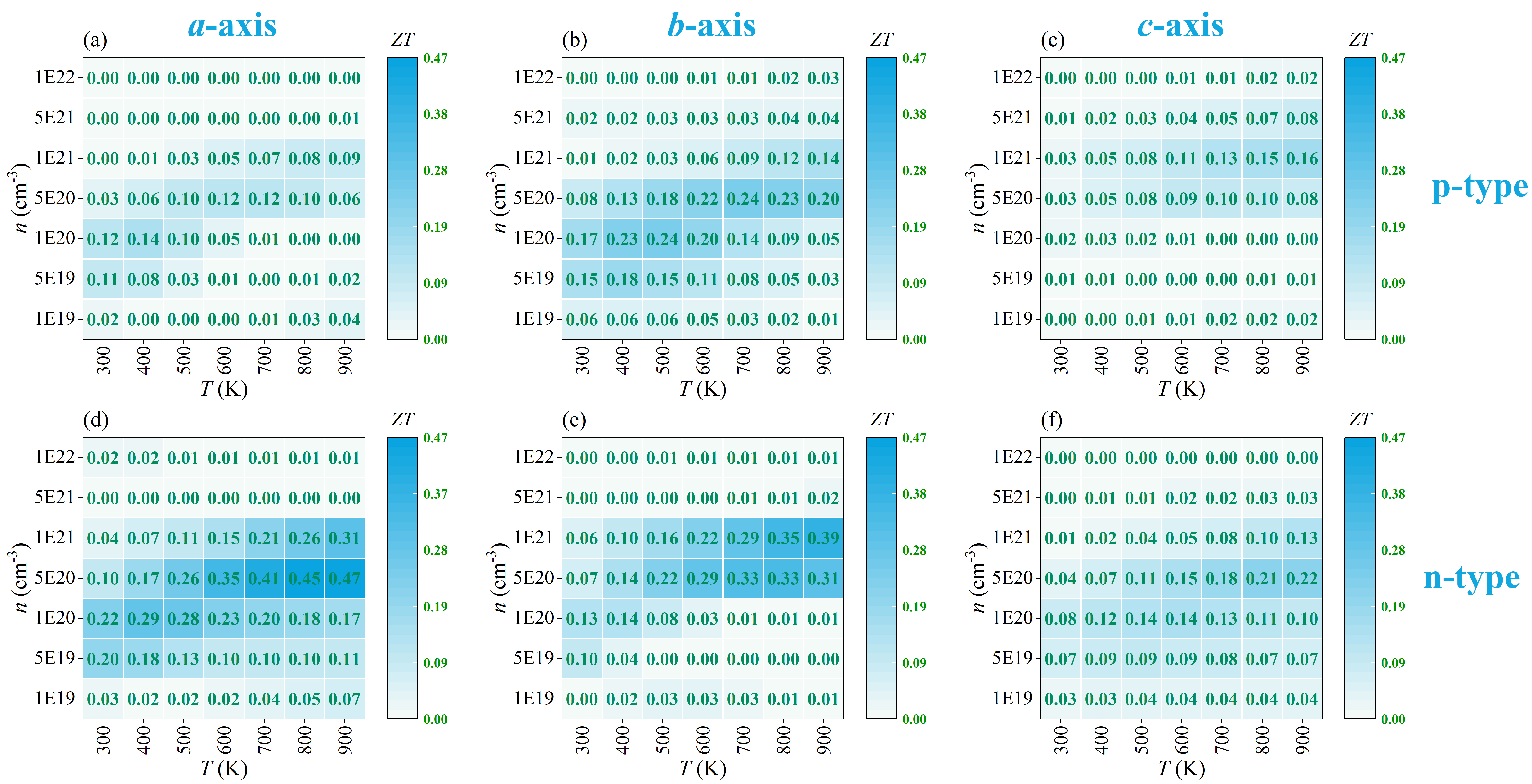}
\caption{(Color online)
The calculated $ZT$ of Ta$_{2}$Pd$_{3}$Te$_{8}$ crystal with respect to carrier density and temperature. The top and bottom panels are for the p- and n-type and the left, middle and right panels are along the $a$, $b$ and $c$ axes, respectively.
    \label{Fig004} }
\end{figure*}

In fact, the Eq.~\ref{ee7} is  derived from the following expression,
\begin{equation}\label{ee8}
\mu_{h(e)}=\frac{q\tau_{h(e)}}{m^c_{h(e)}},
\end{equation}
where $\tau_{h(e)}$ and $m^c_{h(e)}$ are the relaxation time for carriers and the conductivity effective mass (also named by the transport effective mass). Besides, the term $m_{h(e)}^{3/2}$  originates from the contribution of electronic DOS. Therefore, the modified mobility expression can be written as,
\begin{equation}\label{ee9}
  \mu_{e(h)}=\frac{2\sqrt{2\pi}e\hbar^{4}\rho v^2}{3(k_{B}T)^{3/2}{m^{D}_{h(e)}}^{3/2}m^{C}_{h(e)}\lambda_{h(e)}^2},
\end{equation}

In the isotropy crystals with only one VBM/CBM, the band effective mass, the conductivity effective mass and the DOS mass are equal to each other ($m^{B}_{h(e)}=m^{C}_{h(e)}=m^{D}_{h(e)}$). Obviously, combining Eqs.~\ref{ee7} and ~\ref{ee8}  one can obtain the relaxation time expression
\begin{equation}\label{ee10}
  \tau_{e(h)}=\frac{2\sqrt{2\pi}\hbar^{4}\rho v^2}{3(k_{B}T)^{3/2}{m^{D}_{h(e)}}^{3/2}\lambda_{h(e)}^2}.
\end{equation}
The LA velocity $v_\beta$ along the $\beta$ direction in
the single crystal can be approximated as,
\begin{equation}\label{ee11}
v^\beta=\sqrt{\frac{c^\beta}{\rho}},
\end{equation}
where $c^\beta$ is the elastic constant along the $\beta$ direction. Applying a tiny strain along the $\beta$ direction, the corresponding DP constant $\lambda^\beta_{h(e)}$ can be attained \cite{rr34}. Now the anisotropic relaxation time $\tau_{h(e)}^\beta$ is governed by
\begin{equation}\label{ee12}
  \tau_{h(e)}^\beta=\frac{2\sqrt{2\pi}\hbar^{4}c^\beta}{3(k_{B}T)^{3/2}{m^{D}_{h(e)}}^{3/2}{\lambda_{h(e)}^\beta}^2}.
\end{equation}
The above formula can also be applied to the nondegenerate multi-valley semiconductor in which the Fermi level is located in the bandgap and there are two or more VBMs/CBMs. Assuming the VBMs/CBMs located at different positions in
reciprocal space are all equivalent, $m^{D}_{e(h)}$ is the DOS effective mass at one VBM or CBM, which can be approximated as the geometric mean of the effective masses along three principal axes
  \begin{equation}\label{ee13}
 m_{e(h)}^D=\sqrt[3]{m_{11}m_{22}m_{11}}.
 \end{equation}
 It is worth noting that it cannot be written as the total DOS effective mass of all the VBMs/CBMs,
  \begin{equation}\label{ee14}
 m_{h(e)}^{DT}=N_{h(e)}^{\frac{2}{3}}m_{h(e)}^D,
 \end{equation}
 where $N_{h(e)}$ is the number of VBMs(CBMs). Otherwise, this will overestimate inter valley scattering and thus obtain incorrect relaxation time.

TE material is usually a degenerate semiconductor in which the Fermi level crosses the VB or CB leading to many degenerate states. Thus, its $N_{h(e)}$ can be regarded as an infinite value. The transport properties of degenerate semiconductor having metal property are determined by the electronic structure near the Fermi level. Therefore, the $m_{h(e)}^D$ in Eq.(\ref{ee10}) is on behalf of equivalent DOS effective mass in the free electron gas model, which can be obtained by fitting another version of the Mott formula \cite{rr29} for metal or degenerate semiconductor,
\begin{equation}\label{ee15}
 \overline{S}=\frac{2k_B^2Tm_{e(h)}^D}{3q\hbar^2}(\frac{\pi}{3n})^\frac{2}{3}.
 \end{equation}

In this work, the  $m_{h(e)}^D$ value is obtained by fitting the $\overline{S}$ as a function of temperature at a typical carrier density of $5\times10^{20} cm^{-3}$.
 Fig.~\ref{Fig003} shows that the p-type $\overline{S}$  increases with increasing temperature up to 700 K and then decreases. Obviously, the $\overline{S}$  trends of both the p-type and n-type go against Eq. (\ref{ee15}), this is attributed to bipolar effect caused by the small bandgap of about 0.1 eV  in Ta$_{2}$Pd$_{3}$Te$_{8}$ crystal. Thus, the $\overline{S}$ values from 300 K to 500 K are adopted for fitting, which can effectively avoid the influence of the bipolar effect on the $m_{h(e)}^D$ calculation. The $m^D_h$ is 1.99 $m_0$ (electron mass) and slightly heavier than $m^D_e$ of 1.80 $m_0$.
 The maximal $\overline{S}$ values of p- and n-type are 113 $\mu V K^{-1}$ and -119 $\mu V K^{-1}$ (900 K).

 All the calculated $\lambda_{h(e)}^\beta$ values are negative, and $| \lambda_{h}^a |$, $|\lambda_{h}^b|$ and  $|\lambda_{h}^c|$ are 13.64, 14.26 and 12.28 eV, respectively, which are slightly higher than those of the n-type (10.10 eV, 9.77 eV and 10.17 eV). The DP constant reflects the strength of the electron-phonon interaction. Obviously, the hole surfers from stronger phonon scattering due to relatively large DP contants, and the $\lambda_{h}^\beta$ exhibits stronger anisotropy because large difference of DP constants along the \emph{b} and \emph{c} axes is about 2 eV. However, the differences between  $\lambda_{e}^\beta$ values along different direction are lower than 0.5 eV. The $\tau_{h}^a$ and $\tau_{h}^b$ at 300 K are 3.5 and 3.8 fs far smaller than $\tau_{h}^c$ of 7.2 fs. The $\tau_{e}^\beta$ is about two times of $\tau_{h}^\beta$ at the same direction, which is resulting from different the DP condtants, elastic constants and DOS masses. Assuming  $m^c_{h(e)}$ in Eq.~\ref{ee8} equals $m^D_{h(e)}$, one can obtain approximate $\mu_{h(e)}$ values. it is found that the $\mu_{e}^a$, $\mu_{e}^b$ $\mu_{e}^c$ at 300 K are  7.4, 9.3 and 11.9 $cm^{2}V^{-1}s^{-1}$ far larger than the $\mu_{h}$ values along the corresponding direction (3.1, 3.4 and 6.3 $cm^{2}V^{-1}s^{-1}$), and is comparable to the mobility \cite{r16} of Ta$_{2}$Pd$_{3}$Se$_{8}$. These indicate the electrical transport of n-type is superior to that of p-type, and the mobility of Ta$_{2}$Pd$_{3}$Te$_{8}$ needs to be optimized and improved.
   \begin{figure*}[htp]
\centering
\includegraphics[width=1.7\columnwidth]{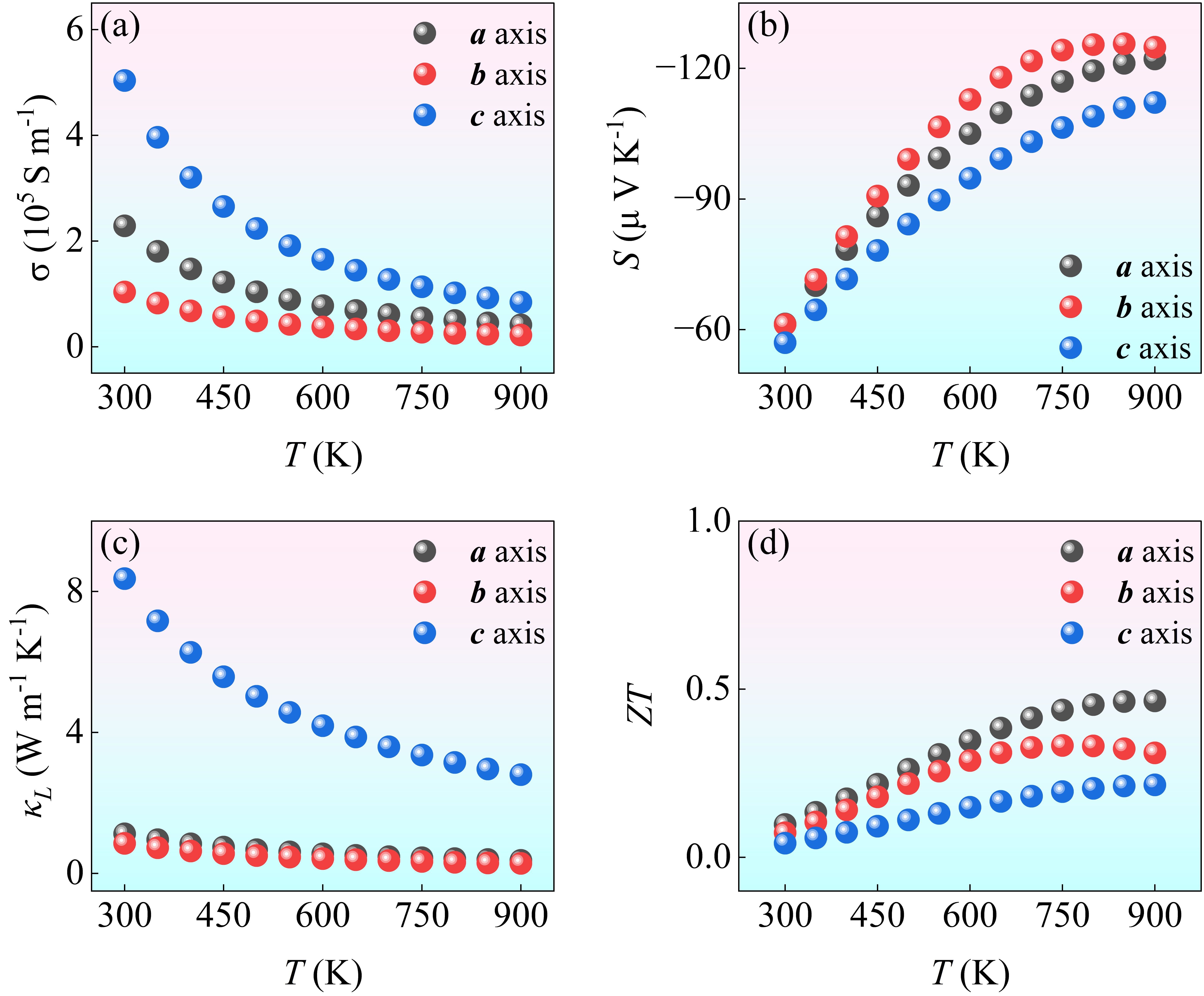}
\caption{(Color online)
TE properties of n-type Ta$_{2}$Pd$_{3}$Te$_{8}$ crystal. The electrical conductivity $\sigma$ (a), Seebeck coefficient $S$ (b), lattice thermal conductivity $\kappa_L$ (c) and the TE figure merit $ZT$ (d) along the $a$, $b$ and $c$ axes as a function of temperature.
 \label{Fig005}
}
\end{figure*}

 After obtaining the relaxation time, all relevant electrical and thermal parameters can be calculated combining the data from the electron Boltzmann equation and phonon Boltzmann equation. Herein, we present $ZT$ along the $a$, $b$ and $c$ directions with respect to carrier density and temperature in Fig.~\ref{Fig004}. For p-type Ta$_{2}$Pd$_{3}$Te$_{8}$ bulk, the optimal $ZT$ values along the $a$, $b$ and $c$ axes (denoted as $ZT^{op}_{11}$, $ZT^{op}_{22}$ and $ZT^{op}_{33}$) are 0.14, 0.24 and 0.16. $ZT^{op}_{11}$, $ZT^{op}_{22}$ and $ZT^{op}_{33}$ appear at 400, 700 and 900 K,  and their corresponding optimal carrier densities are $1\times 10^{20}$, $5\times 10^{20}$ and $1\times 10^{21}$ $cm^{-3}$, respectively. For the n-type, all $ZT^{op}_{11}$, $ZT^{op}_{22}$ and $ZT^{op}_{33}$ appear at 900 K. $ZT^{op}_{11}$ reaches to 0.47 that is higher than $ZT^{op}_{22}$ and $ZT^{op}_{33}$ (0.39 and 0.22). Both $ZT^{op}_{11}$ and $ZT^{op}_{33}$ appear at  $5\times 10^{20} cm^{-3}$, and $ZT^{op}_{22}$ needs a larger carrier density of  $1\times 10^{21} cm^{-3}$. Compared to wide bandgap materials, narrow bandgap materials more easily achieve high carrier density. Moreover, the aforementioned molecular dynamics results indicate Ta$_{2}$Pd$_{3}$Te$_{8}$ crystal is stable at 900 K. The optimal $ZTs$ are very likely to be achieved in the Ta$_{2}$Pd$_{3}$Te$_{8}$ crystal.

  \begin{figure*}[htp]
\centering
\includegraphics[width=1.7\columnwidth]{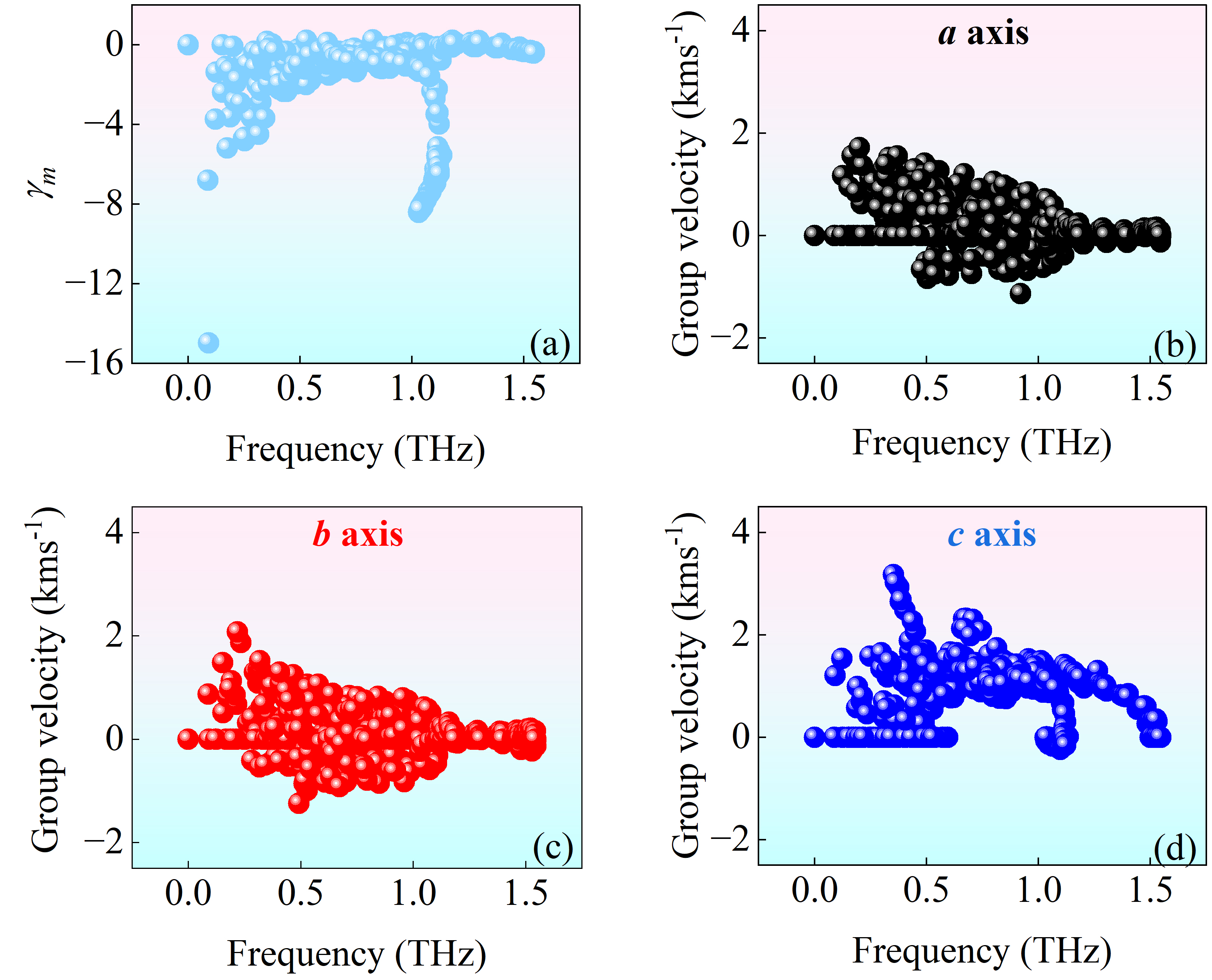}
\caption{(Color online)
Properties of the acoustic branches. The mode Gr\"{u}neisen parameters ($\gamma_m$) as a function of phonon frequency (a). The mode group velocities along the $a$, $b$ and $c$ axes [(b), (c) and (d)] as a function of phonon frequency. Most $\gamma_m$ values are negative, many group velocities along the $a$ and $b$ axes.
    \label{Fig006} }
\end{figure*}

Subsequently, we focus on the TE properties of the n-type Ta$_{2}$Pd$_{3}$Te$_{8}$ because of its relatively high $ZT$ value (see Fig.~\ref{Fig005}), and the p-type was presented in the Supplemental Material Fig. S1. The n-type electrical conductivity in Fig.~\ref{Fig005}(a) exhibits obvious anisotropy near home temperature, and the increasing temperature can suppress the anisotropy.
 The ratio of $\sigma _c$ to  $\sigma_b$ decreases from 2.2 at 300 K to 2.0 at 900 K. The $\sigma _c$ value decreases  from 5.03$\times 10^{20} Sm^{-1} $ at 300 K to 2.28 $\times 10^{20} S m^{-1} $ at 900 K.
 This is because high temperature increases the vibration and the quantity of phonons, thereby enhancing the scattering of electrons suffering from phonons.
 At the same direction and different temperatures, the Seebeck coefficient is negatively correlated with the electrical conductivity.
 At the same temperature, the Seebeck coefficients obey the trend $\lvert S_b \lvert >\lvert S_a \lvert >\lvert S_c \lvert$ that is opposite to that of electrical conductivity.
 Fig.~\ref{Fig005}(b) shows the Seebeck coefficient has weak anisotropy especially at room temperature, and the anisotropy slightly increases with rising temperature. The maximal Seebeck coefficients along $a$, $b$ and $c$ axes reach 122, 125 and 112 $\mu$ K$^{-1}$.

Compared to excellent TE material SnSe, the Seebeck coefficient is lower, however, its electrical conductivity is relatively higher.
Thus, the power factor ($PF$) is superior to that of SnSe. The n-type maximal $PF$ values along $a$, $b$ and $c$ axes reach 9.0, 4.8 and 16.4 $\mu$Wcm$^{-1}$K$^{-2}$ (see the Supplemental Material Fig. S2 (a)), respectively, which are larger than those of the p-type (see the Supplemental Material Fig. S2 (b)).
All the maximal $PF$ values (2.1, 10.1 and 7.7 $\mu$Wcm$^{-1}$K$^{-2}$) of SnSe along the three principal axes appears at high temperature of 850 K.
However, the maximal $PF$ values of Ta$_{2}$Pd$_{3}$Te$_{8}$ appears separately at relatively low temperatures (450, 550 and 350 K), which is due to bipolar effect and is not conducive to improving TE performance.
At 850 K, the $PFs$ of Ta$_{2}$Pd$_{3}$Te$_{8}$ decrease to 6.6, 3.7 and 11.3 $\mu$Wcm$^{-1}$K$^{-2}$. Their average value is comparable to that of SnSe bulk. Unfortunately, the maximal $ZT$ of Ta$_{2}$Pd$_{3}$Te$_{8}$ is only 18\% of that of SnSe. This is mainly due to the total thermal conductivity being much higher than that of SnSe.
As is well known, the total thermal conductivity includes lattice thermal conductivity and electronic thermal conductivity.
In fact, the lattice thermal conductivities of Ta$_{2}$Pd$_{3}$Te$_{8}$ along $a$, $b$ and $c$ axes at 900 K are 0.37, 0.28 and 2.80 W m$^{-1}$ K$^{-1}$ (see Fig.~\ref{Fig005}(c)). $\kappa^l_a$ and $\kappa^l_b$ are slight higher than 0.25 W m$^{-1}$ K$^{-1}$ of SnSe along the $b$ axes where there is an astonishing $ZT$ value of 2.62 \cite{r5}. However, its electronic thermal conductivity is far higher than that of SnSe.
Especially, the $\kappa_a^e$ at 300 K reaches as high as 3.24 W m$^{-1}$ K$^{-1}$ and decreases to 1.61 W m$^{-1}$ K$^{-1}$ at 900 K. Although the $\kappa_b^e$ is as low as 0.73 W m$^{-1}$ K$^{-1}$, it is 2.6 times of corresponding $\kappa_b^l$.
Obviously, the $\kappa_a^e$ and  $\kappa_c^e$ (see the Supplemental Material Fig. S3) generally obey the Wiedemann-Franz law $\kappa^e=L\sigma T$ where $L$ is the Lorenz number. However, the Lorenz number seems to have anisotropy and depend on temperature, because the calculated $\kappa^e$ in the work is gained by the Eq. (18) in Ref. \cite{rr26}, and thus is directly connected with the $PF$. TE properties of p-type and n-type Ta$_2$Pd$_3$Te$_8$ polycrystals are presented in the Supplemental Material Fig. S4, their optimal $ZTs$ only reach to 0.11 and 0.28 due to the absence of anisotropy.

 In a word, high $PF$ and low lattice thermal conductivity are  significant advantages of Ta$_{2}$Pd$_{3}$Te$_{8}$.
 Ordinarily, high $PF$ is accompanied by a byproduct of high electronic thermal conductivity hindering TE performance, and low lattice thermal conductivity caused by strong anharmonic vibration leads to low electrical conductivity.
 However, in Ta$_{2}$Pd$_{3}$Te$_{8}$, low lattice thermal conductivity does not result in low electrical conductivity.
 Therefore, it is necessary and meaningful to analyze the physical and chemical nature for low lattice thermal conductivity.

  \begin{figure*}[htp]
\centering
\includegraphics[width=1.7\columnwidth]{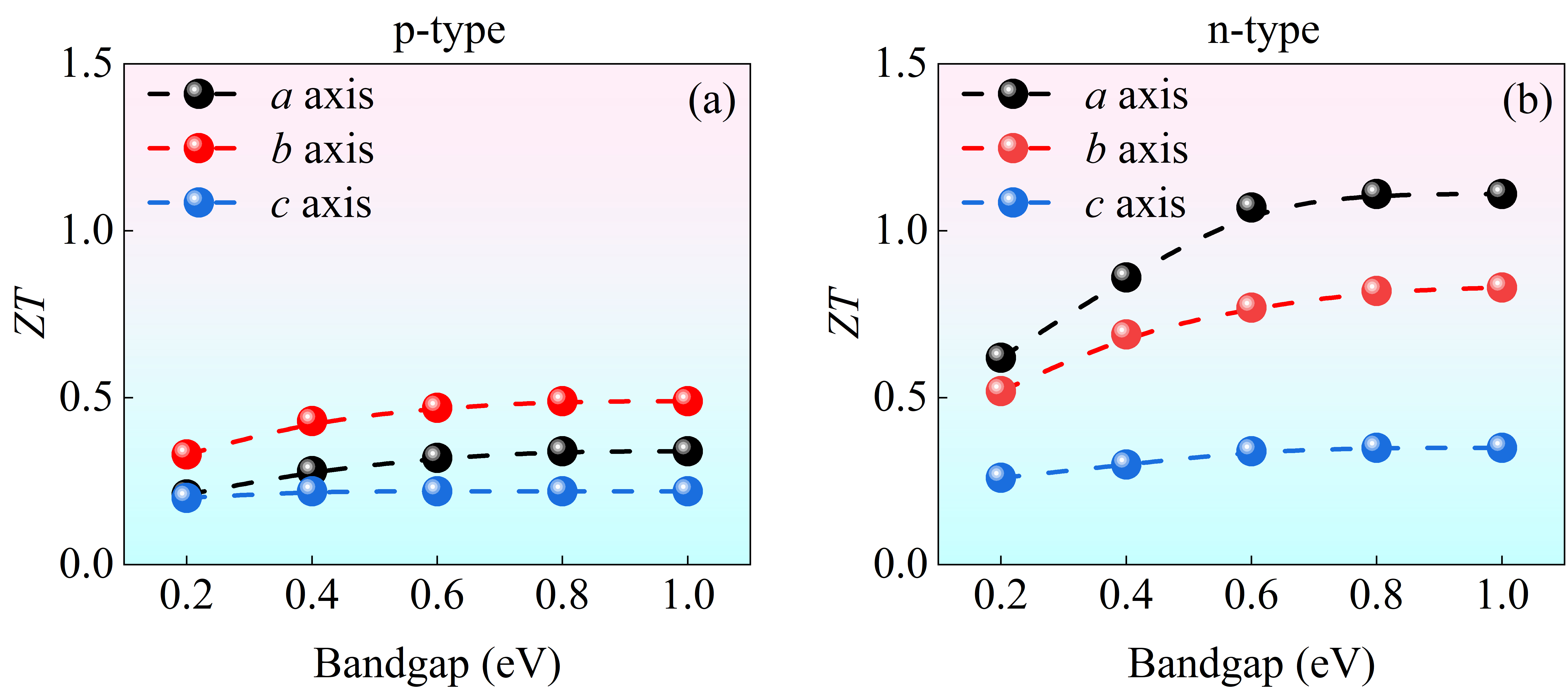}
\caption{(Color online)
The figure merit $ZT$ of p-type (a) and n-type (b) Ta$_{2}$Pd$_{3}$Te$_{8}$ crystal along the $a$, $b$ and $c$ axes as a function of bandgap. $ZT$ along the $a$ axis reaches 1.11 at a bandgap of 0.8 eV.
\label{Fig007} }
\end{figure*}
 According to the kinetic theory of gases, the lattice thermal conductivity is described by
 \begin{equation}\label{ee16}
 \kappa^l=\frac{1}{3}c_V\rho vl,
 \end{equation}
 where $c_V$, $v$ and $l$ stand for the constant-volume specific heat per unit mass, average velocity of phonon, average free path of phonon, respectively. It has two important variants that are usually applied separately for theoretical analysis and experimental measurement,
 \begin{align}\label{ee17}
 &\kappa^l=\frac{1}{3}C_Vvl, \\
 &\kappa^l=c_V \rho D,
 \end{align}
 where $C_V$ and $D$ are the constant-volume specific heat per unit volume and thermal diffusivity coefficient that is given by
 \begin{equation}\label{ee18}
D=\frac{1}{3}vl
 \end{equation}
 It is difficult to obtain $c_V$ experimentally, thus the $c_V$ is replaced by the constant-press specific heat ($c_p$) per unit mass,
  \begin{align}\label{ee19}
 &\kappa^l=c_p \rho D.
 \end{align}
 In the kinetic theory of gas, the influence of velocity distribution on thermal transport is ignored. Obviously, the kinetic theory of gas and the replacing specific heat can lead to a deviation between experimental values and true values of thermal conductivity.
 Therefore, the difference between theoretical and experimental thermal conductivities is not only attributed to theoretical methods, but also to experimental measurements.

 Another commonly used formula \cite{rr30, rr31} for lattice thermal conductivity is proposed by Slack,
 \begin{equation} \label{ee20}
{{\kappa }^{l}}=A\frac{\Theta _{D}^{3}V_{p}^{1/3}\bar{m}}{\gamma _{a}^{2}n_{tot}T}.
\end{equation}

Here, A is a collection related to some physical constants, $\Theta _{D}$ stand for the Debye temperature, $\bar{m}$, $V_{p}$ and $n_{tot}$ are the mean atomic weight,  crystal volume and total number of atoms in the primitive cell, $\gamma _{a}$ denotes the acoustic Gr\"{u}neisen parameter. Note that, when $\bar{m}$, $V_{p}$ and $\kappa^l$ are in units of amu, \AA ~and W/mK, the A value can be given by the expression \cite{rr32}

\begin{equation} \label{ee21}
A=\frac{2.43\times10^{-6}}{1-0.514/\gamma_a+0.228/\gamma_a^2}.
\end{equation}
Slack took $\gamma_a = 2$ in the above expression and obtained $A=3.04\times10^{-6}$. In this work, we obtain $\gamma_a = 1.71$ by the expression \cite{rr33}
\begin{equation}
\gamma_{a}=\frac{9\mathop{v}_{l}^{2}-12\mathop{v}_{t}^{2}}{2\mathop{v}_{l}^{2}+4\mathop{v}_{t}^{2}},
 \end{equation}
 where $v_l$ and $v_t$ stand for the TA and LA velocities that can be computed in virtue of elastic constants.
 Thus, we get $A=3.12\times10^{-6}$, and take $\Theta _{D}$ = 160.9 K obtained  by elastic constants. The $\kappa^l$ at 300 K is 0.65 $W m^{-1}K^{-1}$ that is higher than $\kappa_a^l$ and $\kappa_b^l$ but lower than $\kappa_c^l$.

 The lattice thermal conductivity obtained from Eq. (\ref{ee20}) is based on the assumption of isotropy, and thus it can be regarded as polycrystalline lattice thermal conductivity.
 Obviously, $\kappa^l$ at room temperature is far lower than 3.44 $W m^{-1}K^{-1}$ of ($\kappa^a_L$+$\kappa^b_L$+$\kappa^c_L$)/3 that is ordinarily treated as polycrystalline. Such large difference between $\kappa_L$ possibly reflects the strong and complex anisotropy of Ta$_{2}$Pd$_{3}$Te$_{8}$ crystal.

The anisotropy of crystal  is closely related to the anharmonic vibration of phonons that is the reason for the formation of thermal resistance (reciprocal of thermal conductivity).
The strength of the anharmonic vibration is evaluated by the Gr\"{u}neisen parameter. Most of mode Gr\"{u}neisen parameters ($\gamma_m$) of the three acoustic branches are negative (see Fig.~\ref{Fig006}(a)), and the maximum value of |$\gamma_m$| reach as high as 15, implying strong anharmonic vibration that causes low thermal conductivity. Moreover, we found that anisotropic group velocities leads to anisotropic thermal conductivity.
It is showded in Fig.~\ref{Fig006}(b)-(d) that mode group velocities along the $c$ axis are  significantly greater than those along $a$ and $b$ axes. Most of group velocities along the $a$ axis coincide with those along the $b$ direction.
Additionally, it is worth noting that many mode group velocities along both $a$ and $b$ axes are negative, indicating the lack of coordination in atomic vibration. Therefore, this is highly likely to be related to negative mode  Gr\"{u}neisen parameters.

The inherent advantage of Ta$_{2}$Pd$_{3}$Te$_{8}$  is its low lattice thermal conductivity, which makes it a potential excellent TE material.
How to enhance the TE performance of this material is also the focus of our research. The main shortcoming of this material is strong bipolar effect caused by small bandgap, which results in the maximum $PF$  being in the low temperature zone.
Thus, extending its bandgap to reduce the  bipolar effect is an effective method to improve TE performance.
We assume that the change in bandgap has a negligible impact on the electronic topology structure, and thus can rigidly enlarge the band gap.

Fig. ~\ref{Fig007}(a) and (b) show that the optimal \emph{ZT} increases with increasing bandgap both in p- or n-type bulk, and the optimal figure of merit of p-type bulk is in line with this tendency of $ZT_b > ZT_a > ZT_c$, however, it exhibits $ZT_a > ZT_b > ZT_c$ in n-type bulk. While the bandgap exceeds 0.6 eV, the upward trend of $ZT$ is not obvious and even vanishes. The optimal $ZT_b$, $ZT_a$ and $ZT_c$ of the n-type with a bandgap of 0.8 eV reach 1.11, 0.82 and 0.35, respectively. The optimal $ZT_b$, $ZT_a$ and $ZT_c$ of the p-type have relative low values of 0.34, 0.49 and 0.22. Moreover, the optimal $ZT_c$ of both p- and n-type seems insensitive to bandgap. For instance, the increase in $ZT$ caused by the bandgap change is only 0.02.

\section{CONCLUSION}
In conclusion, we have performed a systematic analysis of stability properties, electronic and phononic structures, and anisotropic TE properties of Ta$_{2}$Pd$_{3}$Te$_{8}$ crystal via the first-principles calculations combined with the Boltzmann transport theory and molecular dynamics
methods. It is found that Ta$_{2}$Pd$_{3}$Te$_{8}$ crystal meets the criteria of mechanical, dynamical, and thermal stabilities well, thus is beneficial for the high-temperature operation of TE devices. Moreover, anisotropic crystal structure leads to anisotropic electrical conductivity, thermal conductivity and Seebeck coefficient and thus provide a good condition for regulating TE performance. We propose the bandgap engineering effectively tunes TE performance. It is predicted the $ZT$ of n-type Ta$_{2}$Pd$_{3}$Te$_{8}$ along $b$ axis can increase from 0.47 to 1.11 by utilizing the variation of bandgap from 0.1 to 0.8 eV.


Aijun Hong: Conceptualization, Data curation, Formal
analysis, Investigation, Writing $–$ original draft, Writing $–$ review $\&$ editing.
Shi Chen: Conceptualization, Resources, Software, Writing $–$ original draft,
Writing $–$ review $\&$ editing.
Junming Liu: Conceptualization, Data curation, Methodology, Resources, Software,
Writing $–$ original draft, Writing $–$ review $\&$ editing.\\

\noindent{\textbf{Declaration of competing interest}}\\

The authors declare that they have no known competing financial interests or personal relationships that could have appeared to influence the work reported in this paper.\\

\noindent{\textbf{Data availability}}\\

No data was used for the research described in the article.\\

\noindent{\textbf{Acknowledgement}}\\

This work was supported by the National Natural Science
Foundation of China (Grant No. 11804132).
\\
~

\bibliographystyle{aapmrev4-2}
\bibliography{apsguide4-2}
\end{document}